\documentclass[twocolumn,superscriptaddress,floatfix,preprintnumbers,prl,hyperref,nofootinbib,nobibnotes,amsmath,amssymb,aps]{revtex4-1}

\usepackage{graphicx}
\usepackage{dcolumn}
\usepackage{bm}
\usepackage[mathlines]{lineno}
\usepackage{color}
\usepackage{xcolor}
\usepackage{verbatim}
\usepackage{soul}

\begin{document}

\preprint{FERMILAB-PUB-24-0576-T,~~ PITT-PACC-2405}
\title{Electroweak Symmetry Restoration and Radiation Amplitude Zeros}
\author{Rodolfo Capdevilla}
\affiliation{Particle Physics Department, Fermi National Accelerator Laboratory, Batavia, IL 60510, USA}
\author{Tao Han}
\affiliation{Department of Physics and Astronomy, University of Pittsburgh, Pittsburgh, PA 15217, USA}
\date{\today}

\begin{abstract}
In high-energy collisions far above the electroweak scale, the effects of electroweak symmetry breaking are expected to become parametrically small $\delta \sim M_W/E$. This defines the extent to which the electroweak gauge symmetry is restored: $(i)$ the physics of the transverse gauge bosons and fermions is described by a massless theory in the unbroken phase; $(ii)$ the longitudinal gauge bosons behave like the Goldstone bosons and join the Higgs boson to restore the unbroken $O(4)$ symmetry in the original Higgs sector. Using the unique feature of the radiation amplitude zeros in gauge theory, we propose to study the electroweak symmetry restoration quantitatively by examining the processes for the gauge boson pair production $W^\pm \gamma,\ W^\pm Z$ and $W^\pm H$ at the LHC and muon colliders. 
\end{abstract}

\maketitle


{\bf Introduction} - The discovery of the Higgs boson completes the particle spectrum for the Standard Model (SM) of elementary particle physics as a self-consistent theory potentially valid to an exponentially high scale. Yet, the exploration for physics beyond the electroweak scale continues to drive the energy frontier to seek for new physics beyond the SM (BSM) and to appreciate the rich physics on its own right. 

In reaching the 10 TeV partonic center-of-momentum (c.m.)~energy, one would expect to enter a qualitatively new regime. Comparing to the scale ($v$) of the electroweak symmetry breaking (EWSB), $v/10$ TeV$\approx 2.5\times 10^{-2}$, one starts to entertain the notion of the ``electroweak symmetry restoration" (EWSR). Indeed, it would be analogous to QCD physics at the scale of 10 GeV,  $\Lambda_{QCD}/10$ GeV$\approx 2.5\times 10^{-2}$, where the physics enters its symmetric phase described by massless quarks ($u,d,s$) and gluons, rather than hadrons in the broken phase. It is thus appropriate to define the EWSR as that the physics is governed in its symmetric phase with massless particles as gauge multiplets and their interactions dictated by its full gauge symmetry.  

What first comes to mind when thinking about EWSR is the Goldstone boson Equivalence Theorem (GET) \cite{Cornwall:1974km,Lee:1977eg,Chanowitz:1985hj}. It states that the scattering amplitudes of longitudinal gauge bosons at high energies are equivalent to those of their corresponding Goldstone bosons. For an on-shell massive vector boson $p^2=M^2$, the longitudinal polarization vector can be written as\footnote{For a on-shell massive vector particle, this form is dictated by its Lorentz properties, regardless of its gauge nature.}  
\begin{equation}
\epsilon^\mu_L(p) = {E\over M}(\beta, {\hat p}) = {p^\mu\over M} - {1\over {1+\beta}}{M\over E}n^\mu, 
\end{equation}
where $\beta=p/E$ is the speed and $n^\mu=(1, -\hat p)$ is a light-like four-vector. Contracting it to obtain the physical matrix element, the first momentum term ``scalarizes'' the amplitude, in accordance with GET at high energies $E\gg M$ (or $\beta\to 1)$. The second term measures the symmetry breaking effect and thus the deviation from the GET. Thus, we define
\begin{equation}
\delta \equiv {1\over 2}{M\over E} 
\label{eq:d}
\end{equation}
to quantify the residual effect of the EWSB.  

We stress that the Goldstone bosons only specify the broken symmetry \cite{Goldstone:1962es} as a subset of a higher-scalar representation in a UV complete theory. In the SM, the three Goldstone bosons ($\omega^{\pm,0}$) form an SU(2) custodial triplet \cite{Sikivie:1980hm} corresponding to the broken generators in SU(2)$_L\otimes$U(1)$_Y\to$U(1)$_{\rm em}$. Beyond the SM in terms of an Effective Field Theory (EFT) \cite{Alonso:2015fsp,Falkowski:2019tft,Cohen:2020xca}, their representation corresponds to the Higgs Effective Field Theory (HEFT) as the nonlinear realization of the gauge symmetry, in contrast to the SM Effective Field Theory (SMEFT) as the linear realization. To observe EWSR for the full symmetry in the SM, we advocate the conditions:  
\begin{itemize}
    \item[$(i)$] the physics of the transverse gauge bosons $(W^\pm_T, Z_T, \gamma)$ and fermions is described by a massless theory in the unbroken phase;
    \item[$(ii)$] the longitudinal gauge bosons $(W^\pm_L Z_L)$ are scalarized as Goldstone bosons $(\omega^\pm,\omega^0)$, and join the Higgs boson to restore the unbroken $O(4)$ symmetry $(\omega^\pm,\omega^0,H)$ in the Higgs sector.
\end{itemize}
Both conditions above can be quantitatively measured by the residual EWSB effect $\delta$, and should be applicable to any spontaneously broken theories. In particular, condition $(ii)$ would provide a quantitative assessment of the Higgs sector, potentially sensitive to BSM physics, such as 
SMEFT with a Higgs doublet versus HEFT with a non-Standard Model-like Higgs singlet, or even revealing an extended Higgs sector with a larger global symmetry 
\cite{Branco:2011iw}.

\begin{figure*}
\centering
\includegraphics[width=0.4\linewidth]{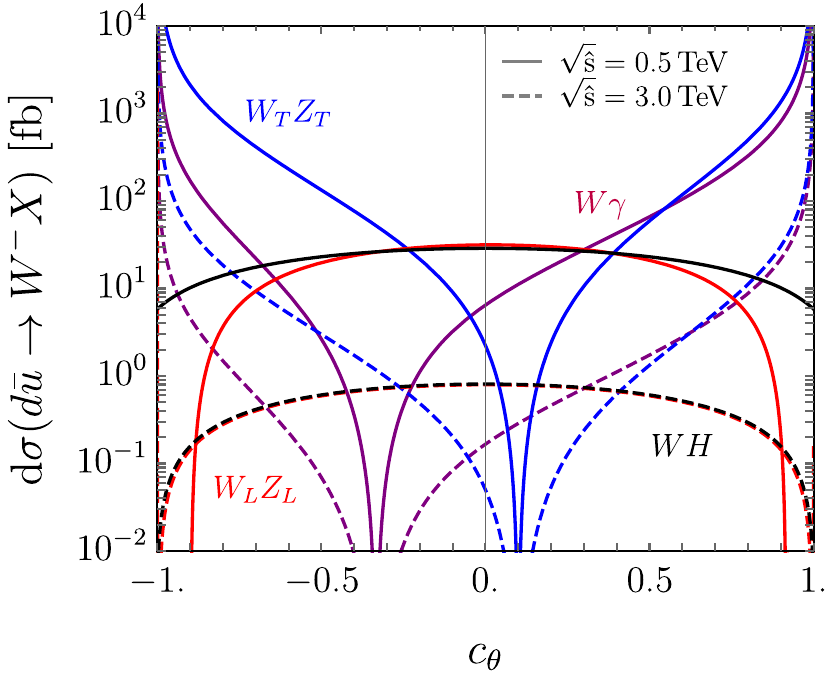}~~~~~~~~\includegraphics[width=0.4\linewidth]{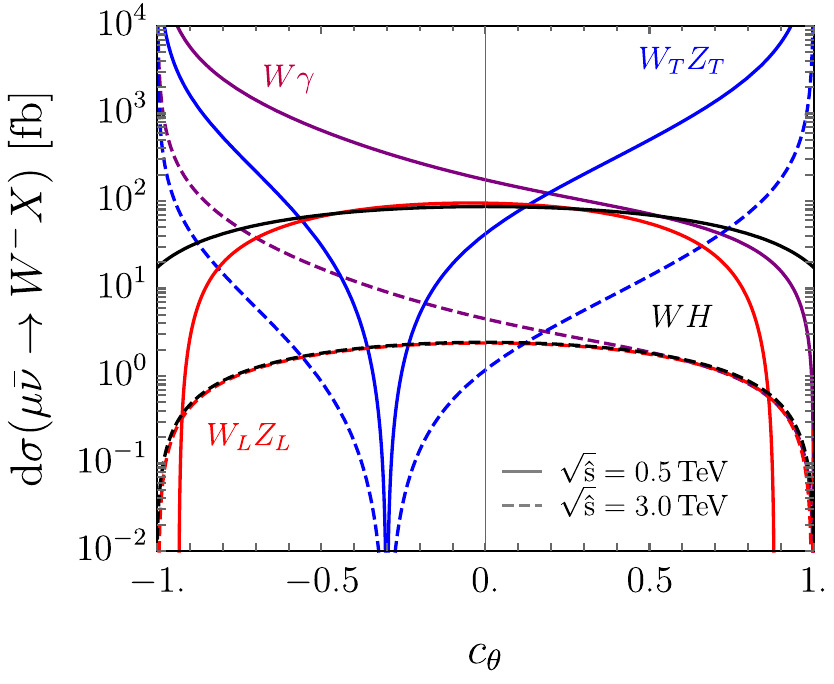}
\caption{
Angular distributions of processes $d\bar u \to W^-X$ (left) and $\mu^- \bar \nu \to W^-X$ (right), where $X=\gamma,Z,H$, at a c.m. energy of 0.5 TeV (solid) and 3 TeV (dashed). Both $W\gamma$ (purple) and the transverse modes $W_TZ_T$ (blue) depict a RAZ. The longitudinal modes $W_LZ_L$ (red) and $WH$ (black-dashed) are symmetric, and exactly overlap at high energies. 
}
\label{fig:RAZ}
\end{figure*}

The gauge structure of the SM has been established and tested to high precision by a large number of experiments in recent decades \cite{ParticleDataGroup:2024cfk}. The H1 and ZEUS collaborations in HERA with $e^\pm p$ collisions at $E\sim 100$ GeV first established the unification of neutral currents (via $\gamma/Z$) and charged currents (via $W^\pm$) \cite{H1:2015ubc}. At LEP 2, the measurements for the forward-backward asymmetry reached the precision to test the unbroken SU(2)$_L\otimes$U(1)$_Y$ gauge interaction to an accuracy better than $M_Z^2/s \sim 20\%$ \cite{Peskin:2017emn}. The longitudinal behavior of the gauge boson first manifested itself in the top-quark decay $\Gamma(t\to b W_L)/\Gamma(t\to b W_T) \approx  m_t^2/2M_W^2 \approx 2$ \cite{Kuhn:1996ug,CMS:2020ezf}. The milestone discovery of the Higgs boson \cite{ATLAS:2012yve,CMS:2012qbp}  provides crucial evidence for a SM particle spectrum as a UV-complete gauge theory. Recent results of gauge boson pair production at the LHC confirmed the existence of longitudinal gauge boson contributions with a high statistical significance \cite{ATLAS:2022oge, ATLAS:2024qbd}.
The current studies for $WW$ scattering at the LHC \cite{CMS:2020etf,ATLAS:2019cbr} are mostly sensitive to transverse gauge-boson interactions. 
The searches of same-sign $W$ boson pairs in association with two jets found a $3.3\,\sigma$ significance in the production of at least one longitudinally polarized $W$ boson \cite{ATLAS:2025wuw}.
Thus far, all measurements are near the EW scale $E\approx v$ in a broken phase measured by $\delta \approx M_W/E \approx 0.3$. The authors of \cite{Huang:2020iya} proposed to establish the equivalence between the longitudinal gauge boson and the Goldstone boson $\sigma(W_L H)/ \sigma(\omega H)\to 1$ in a proper limit $v \to 0$ towards the observation of EWSR. Ultimately, the longitudinal gauge boson scattering at high energies \cite{Lee:1977eg,Chanowitz:1985hj} would reveal the full structure of the scalar sector in the symmetric phase. 

This Letter attempts to address EWSR quantitatively at colliders. We propose to carry out a comparative study for the processes of gauge boson pair production.
\begin{eqnarray}
\label{eq:wgm_0}
&& f_1 \bar f_2 \to W^\pm \gamma , \\
\label{eq:wz_0}
&& f_1 \bar f_2 \to W^\pm Z, \\ 
\label{eq:wh_0}
&& f_1 \bar f_2 \to W^\pm H .
\end{eqnarray}
It was long realized that the $W\gamma$ process Eq.~(\ref{eq:wgm_0}) exhibits a peculiar radiation amplitude zero (RAZ) at a particular angle specified by the electric charges of the particles involved~\cite{Mikaelian:1979nr,Brodsky:1982sh}. In addition, it was later realized that the $WZ$ process of Eq.~(\ref{eq:wz_0}) also leads to an approximate RAZ~\cite{Baur:1994ia}. The important observation made there is that while the transverse gauge bosons do exhibit an exact RAZ governed by the electroweak gauge charges, the Goldstones and the EWSB effects tend to
fill in the zeros. They are thus ideal processes for effectively studying the EWSR conditions with the following expectations:
\begin{itemize}
    \item[(a)] Transverse gauge boson production $W^\pm_T \gamma$ and $W^\pm_T Z_T$ exhibit RAZs at high energies described by a massless gauge sector.
    \item[(b)] Longitudinal gauge boson production $W^\pm_L Z_L$ is symmetric with no RAZ, behaves similarly to Goldstone bosons, and approaches the Higgs counterpart $W^\pm_L H$, restoring the original $O(4)$ multiplet $(\omega^\pm,\omega^0,H)$ in the Higgs sector~\cite{Lee:1977eg},
\end{itemize}
again, both quantitatively measured by Eq.~(\ref{eq:d}).

In the remainder of this Letter, we first discuss the characteristic features of the RAZs in $W\gamma/WZ$ processes and define quantitative observables for EWSR. We then propose the search strategy for observing EWSR at the High Luminosity LHC (HL-LHC) and at a high-energy muon collider (MuC). Finally, we provide some further remarks on EWSR and summarize our results. 


{\bf Radiation Amplitude Zeroes} - We denote the tree-level helicity amplitude for $W^-X$ production as ${\cal M}^{WX}_{\lambda_{\rm w}\lambda_X}$, where the polarization of particle $i$ can be either $\lambda_i=\pm$ (transverse) or $\lambda_i=0$ (longitudinal). At high energies, $\sqrt{s} \gg M_W$, the amplitudes of the processes in Eqs.~(\ref{eq:wgm_0})$-$(\ref{eq:wh_0}) can be expressed in terms of the polar angle $\theta\ (c_{\theta}=\cos\theta)$ of \( W \) with respect to \( f_1 \) in the partonic c.~m.~frame.
\begin{eqnarray}
\label{eq:wgm}
{\cal M}^{W\gamma}_{\pm\mp} &\approx& -{ge V_{12} \over \sqrt 2} {(\lambda_{\rm w} - c_{\theta }) \over s_{\theta}} \Bigl[ Q_{(1-2)} c_{\theta} - Q_{(1+2)} \Bigr],\quad\quad \\ 
{\cal M}^{WZ}_{\pm\mp} &\approx& {g g_z V_{12} \over \sqrt 2} {(\lambda_{\rm w} - c_{\theta}) \over s_{\theta }} \Bigl[  g_-^{(1-2)} c_{\theta }   - g_-^{(1+2)} \Bigr],\label{eq:wzT} \\
{\cal M}^{WZ}_{00} &\approx& -{ g_z^2 V_{12} \over 2\sqrt 2} s_{\theta } g_-^{(1-2)}={ g^2 V_{12} \over 2\sqrt 2} s_{\theta },\label{eq:wzL} \\ 
{\cal M}^{WH}_0 &\approx& { g^2 V_{12} \over 2\sqrt 2} s_{\theta },
\label{eq:wh}
\end{eqnarray}
where, \( e = g \sin \theta_w \), \( g_z = g / \cos \theta_w \), \(\theta_w\) is the weak mixing angle, \( V_{12} \) is the flavor mixing element, \( Q_{(1\pm2)} = Q_1 \pm Q_2 \), \( g_-^{(1\pm2)} = g_-^{f_1} \pm g_-^{f_2} \) with \( g^{f_i}_- = T_3^i - \sin^2 \theta_w Q_i \), and \( T_3^i \) (\( Q_i \)) the weak isospin (electric charge) of the fermion \( f_i \).  

Following Eqs.~(\ref{eq:wgm})$-$(\ref{eq:wh}), the RAZs manifest themselves for the transverse gauge bosons ($T=\pm$) at the locations 
\begin{eqnarray}
c_{\theta_0} & = &\hskip-0.2cm  
\left\{  
\begin{array}{ll}
-1/3\ (\approx 0.1) &   {\rm  for}\ d\bar u\to W^-_T\gamma\ (W^-_T Z_T), \\
1\ (\approx -0.3)  &   {\rm for}\ \ell^-\bar \nu \to W^-_T\gamma\ (W^-_T Z_T), ~~~~~  
\end{array}
\right.
\label{eq:raz}
\end{eqnarray}
whereas longitudinal gauge bosons $W_L Z_L$ ($L=0$) fill in the zero region and approach the Higgs boson production. The characteristic angular distributions of processes (\ref{eq:wgm_0})$-$(\ref{eq:wh_0}) are shown in Fig.~\ref{fig:RAZ}  for $\sqrt {\hat s} =0.5$ TeV (solid) and 3 TeV (dashed). The $W\gamma$ (purple) and $W_TZ_T$ (blue) processes show the RAZs, respectively. The $W_LZ_L$ (red) and $WH$ (black) processes, on the other hand, behave similarly and fill in the RAZs. Note that they completely overlap at 3 TeV. Other channels, such as $W_TZ_L$ and $W_LZ_T$, present the residual effect from EWSB, which is parametrically suppressed in terms of $\delta$. 

These features offer a unique opportunity to examine the properties of the gauge bosons and the scalars at high energies separately. Following conditions $(a)$ and $(b)$, we therefore propose to quantify the EWSR by examining the two cross section ratios.
\begin{equation}
    r_{Z\gamma}^{} = { \sigma(W Z) \over \sigma(W \gamma) }\ ,\quad 
    r_{ZH}^{} ={ \sigma(W Z) \over \sigma(W  H)}\ .
    \label{eq:r}
\end{equation}
The cross sections are largely dominated by the transversely polarized gauge bosons. We find that the leading-forward scatterings give the simple behavior 
\begin{equation}
    r_{Z\gamma}^{} \approx { \sigma(W_T Z_T) \over \sigma(W_T \gamma) } \approx {g_z^2\over e^2} 
    { { (g_-^{f_1})^2 + (g_-^{f_2})^2} \over {Q_1^2 +Q_2^2} }\ .
    \label{eq:zr}
\end{equation}
This ratio shows similar features of $Z_T$ and $\gamma$, which only differ by gauge charges, confirming condition (a). As expected, it has little dependence on the collision energy and approaches $r_{Z\gamma}^{} \approx 3.1 \ (1.8)$ for the $d\bar u\ (\mu^-\bar\nu)$ process, as predicted in Eq.~(\ref{eq:zr}). It is interesting to note that if the gauge couplings SU(2)$_L$ and U(1)$_Y$ had the same strength $g=g'$ (or $\sin \theta_w =\cos \theta_w$), then the ratio would be unity $r_{Z\gamma}^{} = 1$. 

\begin{figure}[t]
\centering
\includegraphics[width=0.9\linewidth]{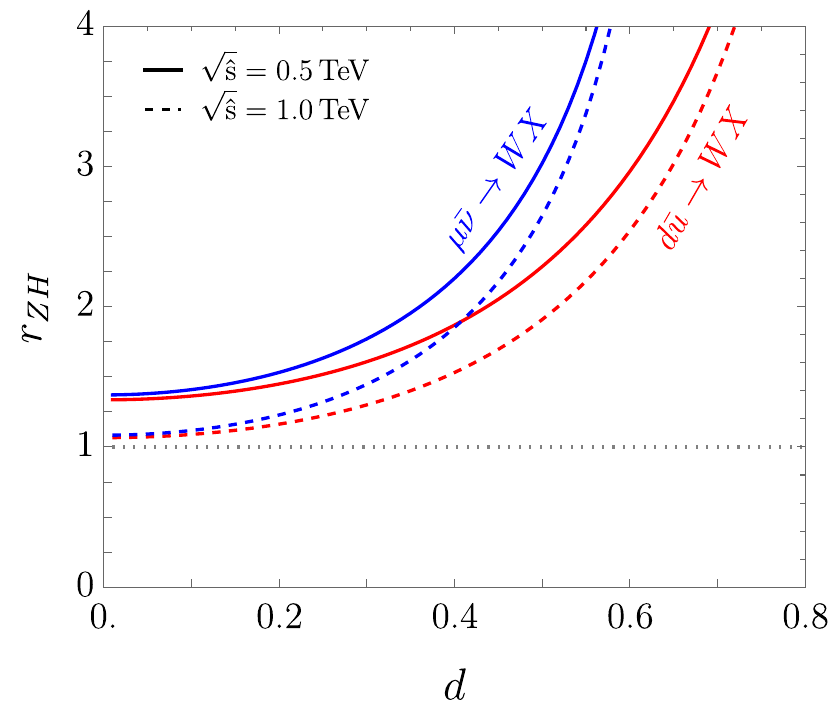}
\caption{
Cross section ratio $r_{ZH}^{}$ defined in Eq.~(\ref{eq:r}) as a function of the angular cut $\Delta c_\theta=c_{\theta_0}\pm d$, where $c_{\theta_0}$ is the location of the RAZ provided in Eq.~(\ref{eq:raz}). As the angular cut gets narrower, the ratio $r_{ZH}^{}$ converges to 1 at high energies. 
}
\label{fig:ratio}
\end{figure}

In contrast, $r_{ZH}^{}$ features the scalar sector. At high energies, in accordance with Eqs.~(\ref{eq:wzL}) and (\ref{eq:wh}), we expect $\sigma(W_L^\pm Z_L) \sim \sigma(W_L^\pm H)$, or $\sigma(\omega^\pm \omega^0)\sim \sigma(\omega^\pm H)$ in lieu of GET, quantitatively establishing  the $O(4)$ symmetry among the multiplet $(\omega^\pm,\omega^0,H)$ in the Higgs sector. In reality, however, $r_{ZH}^{}$ in Eq.~(\ref{eq:r}) is again dominated by transverse gauge bosons $(W_TZ_T)$. In order to reduce transverse contamination, we therefore introduce a selection cut $\Delta c_\theta = c_{{\theta_0}}\pm d$, to focus on the RAZ region guided by Eq.~(\ref{eq:raz}). We show in Fig.~\ref{fig:ratio} how $r_{ZH}\to 1$ around the RAZ region for $\sqrt s = 500$ GeV (solid curves) and 1 TeV (dashed curves). We argue that this convergence at high energies as $\delta \to 0$ signals the restoration of the $O(4)$ symmetry in the Higgs sector. This provides a quantitative demonstration of condition $(b)$ above. 

\begin{figure*}[t]
\centering
\includegraphics[width=0.45\linewidth]{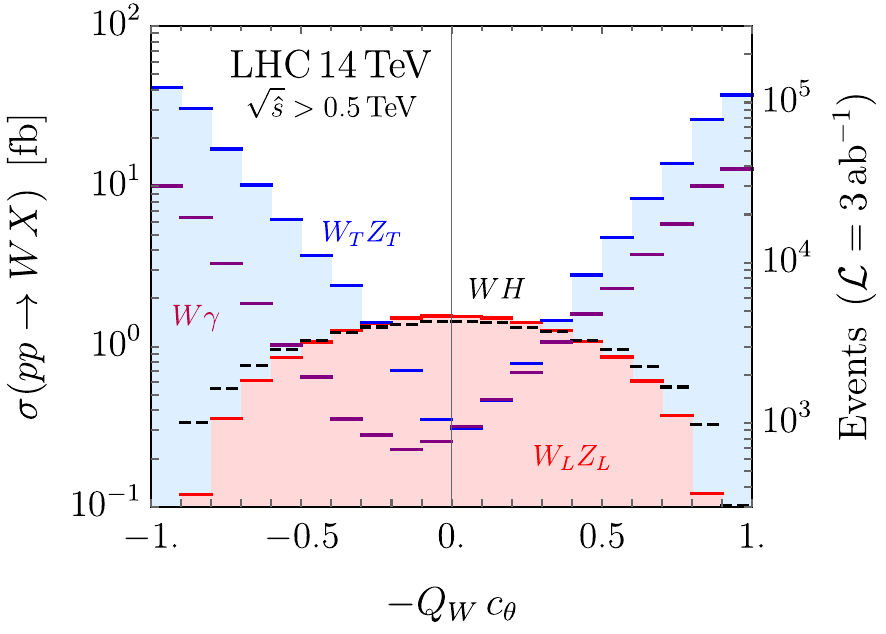}~~~~~~~~\includegraphics[width=0.45\linewidth]{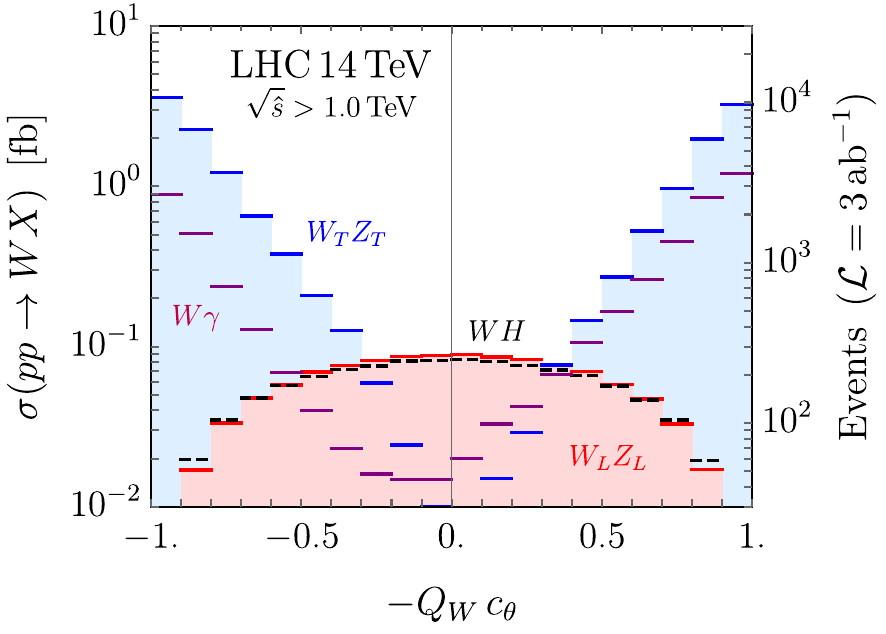}
\caption{
Angular distributions for $p p \to W^\pm X$, where $X= \gamma, Z, H$ at the 14 TeV LHC for a minimum invariant mass of the final state bosons of $M_{WX}=\sqrt{\hat s}>500$ GeV (left) and $M_{WX}>1$ TeV (right). The horizontal axis is the cosine of the $W$ polar angle with respect to the boost of the final state multiplied by the $W$ boson electric charge $(Q_W)$. The vertical right axis show the number of events per bin corresponding to an integrated luminosity of 3 ab$^{-1}$.
}
\label{fig:LHC}
\end{figure*}

\begin{figure*}[t]
\centering
\includegraphics[width=0.45\linewidth]{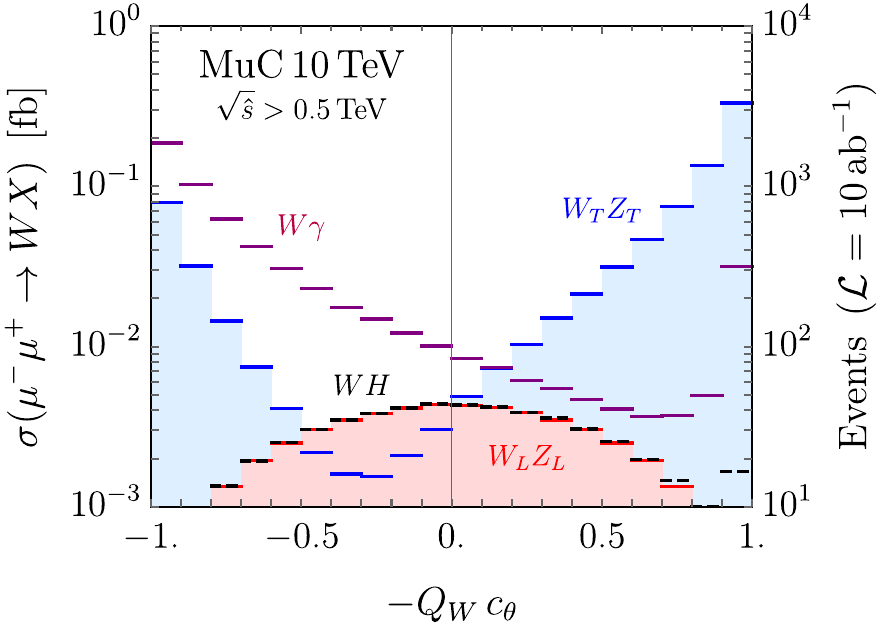}~~~~~~~~\includegraphics[width=0.45\linewidth]{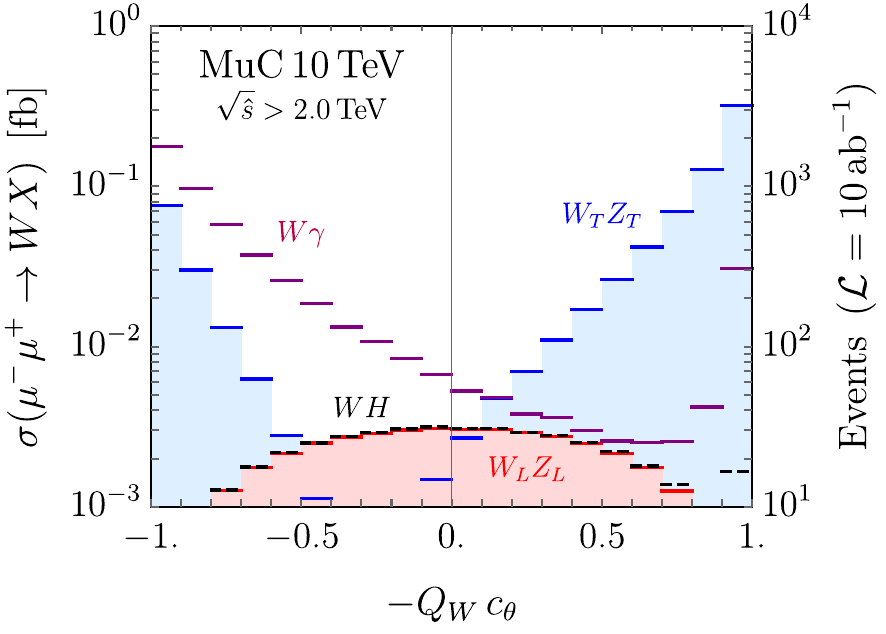}
\caption{
Angular distributions for $\mu^+ \mu^-  \to W^\pm X$ at a 10 TeV MuC for $M_{WX}>500$ GeV (left) and $M_{WX}>2$ TeV (right). The vertical right axis show the number of events per bin corresponding to an integrated luminosity of 10 ab$^{-1}$. 
}
\label{fig:muon}
\end{figure*}


{\bf EWSR at the LHC} - Experimental efforts to observe the RAZ can be traced back to the Fermilab Tevatron \cite{D0:2008abl}. More recently, CMS has observed the RAZs in \( W\gamma \) \cite{CMS:2013ryd,CMS:2021cxr} and ATLAS achieved an impressive measurement of the RAZ in the \( WZ \) final state  in the presence of $W_LZ_L$ \cite{ATLAS:2024qbd}. The associate production channels $WH/ZH$ have also been established at the LHC \cite{ATLAS:2020fcp,CMS:2023vzh}.

To observe the EWSR effects described above, we present the angular distributions for \( pp \to W^\pm X \), where \( X = \gamma, Z, H \), as shown in Fig.~\ref{fig:LHC}. The x-axis displays the cosine of the angle of the \( W \) boson relative to the boost direction of the final state in the partonic c.m.~frame. Since the RAZ is symmetric with respect to \( c_\theta = 0 \), we combine \( W^\pm \) events by multiplying the corresponding \( W \) boson electric charge $Q_W$ on $c_\theta$. We generate events using MadGraph5\_aMC@NLO~\cite{Alwall:2014hca} with specific polarizations of the vector bosons as described in \cite{Ruiz:2021tdt}. (Our event selection includes the kinematic cuts on the photon  $p_T^\gamma>10$ GeV and $|\eta_\gamma|<2.5$).
To investigate the RAZ at high energies, we compare the angular distribution for two acceptance cuts in the invariant mass of the final-state bosons: \( M_{WX} \geq 500 \) GeV (left) and \( M_{WX} \geq 1 \) TeV (right) in Fig.~\ref{fig:LHC}. As expected, the RAZ becomes more pronounced at higher energies, as indicated by the depth of the blue histogram near the $WZ$ zero; the longitudinal modes dominate in this region (red), and contamination from other polarizations is parametrically suppressed by $\delta$. The vertical axes on the right side show the significant number of events per bin expected at the HL-LHC.

It is challenging to measure a ``zero.'' A variety of effects can wash out the RAZ at hadron colliders. This includes detector resolution effects~\cite{Valenzuela:1985dp}. In the most recent measurement of the RAZ~\cite{CMS:2021cxr}, effects such as photon ID, $j/e\to\gamma$ misidentification, and lepton misidentification are the largest sources of systematic uncertainty on expected yield. Another important effect that can wash out the RAZ is higher-order QCD corrections, including final-state radiation~\cite{Monyonko:1984xf} and 1-loop contributions~\cite{Laursen:1982iv,Laursen:1983kw,Smith:1989xz,Ohnemus:1993wx}. Although the RAZ only manifests itself at tree-level, authors in~\cite{Baur:1993ir,Baur:1994sa} demonstrated that by applying a jet veto and measuring rapidity differences, it is possible to visualize the RAZ at hadron colliders (see also~\cite{Capdevilla:2019zbx}). Despite the aforementioned challenges, the measurement of the RAZ in production \( W\gamma \) has been established both at the Tevatron and the LHC~\cite{D0:2008abl,CMS:2013ryd,CMS:2021cxr}.
Likewise, the measurement of $WZ$ production has been established with high confidence~\cite{ATLAS:2024qbd}, including the observation of the RAZ central angular region in the $c_\theta$ distribution where the longitudinal modes dominate.

The most challenging final state to measure our proposed observables of Eq.~(\ref{eq:r}) is $WH$ as it  relies on Higgs tagging. The LHC Run~2 has measured Higgs boson production in association of a leptonically decaying vector boson~\cite{ATLAS:2020fcp,CMS:2023vzh}. 
The projected improvements go well above $5.0\,\sigma$ \cite{CMS:2018qgz}. 
These projections suggest that it is promising to begin to scrutinize the EWSR at the level of $\delta\approx M_W/1~{\rm TeV} < 10\%$ at the HL-LHC. Techniques to improve longitudinal gauge boson tagging are also instrumental in performing our proposed measurements~\cite{Han:2009em,Searcy:2015apa,Kim:2021gtv,Dao:2023pkl}.

{\bf EWSR at a Muon Collider} - A multi-TeV Muon collider has reemerged in recent years as a potential candidate for physics at the future energy frontier~\cite{Long:2020wfp,AlAli:2021let,Black:2022cth,Accettura:2023ked}. The copious production of Higgs bosons via vector boson fusion marks a promising avenue for a precision Higgs program~\cite{Han:2020pif,Forslund:2022xjq,Ruhdorfer:2023uea,Forslund:2023reu,Andreetto:2024rra,Li:2024joa}, complementing the efforts of a dedicated Higgs factory. Furthermore, the potential to reach 10 TeV c.~m.~collisions offers the opportunity to probe highly motivated theories of BSM physics~\cite{Han:2020uak,Buttazzo:2020uzc,Capdevilla:2021fmj,Bottaro:2021snn,Chen:2022msz,Bottaro:2022one,Liu:2023jta,Korshynska:2024suh,Capdevilla:2024bwt}. At energies well above the electroweak scale, all SM particles are relatively light and may be treated as partons of the high-energy incoming muons \cite{Han:2020uid,Han:2021kes,Garosi:2023bvq}. Of particular interest is the collision $\mu^\pm \nu \to W^\pm X$, where the $\nu$ beam is from the soft $W$ radiation $\mu\to \nu W$, effectively producing a high energy $\mu\nu$ collider \cite{Han:2020uid,Capdevilla:2024ydp}. As shown in Fig.~\ref{fig:RAZ} for these unique processes \cite{Baur:1994ia}, the transverse gauge bosons manifest the RAZs, and the longitudinal gauge boson and Higgs exhibit the scalar distributions. 

Using MadGraph5\_aMC@NLO, we carry out a full tree-level simulation for $\mu^+\mu^-\to W^+ W^- X$. We assume that the detector coverage is up to $10^\circ$ ($|\eta|<2.44$). We demand that one of the $W$ bosons be a collinear radiation (with $|\eta|>2.44$), and the other two bosons $W^\mp X$ fall inside the detector and are essentially back-to-back in the transverse plane. Our calculation captures the leading kinematic characteristic of $\mu^\pm \nu$ collisions. We then compute the cosine of the angle of $W^\mp$ in the $W^\mp X$ c.~m.~frame. We show our results in Fig.~\ref{fig:muon} at 10 TeV MuC with \( M_{WX} \geq 500 \) GeV (left) and \( M_{WX} \geq 2 \) TeV (right). The predicted distributions show the RAZ at $c_\theta \approx -0.3$ for $W^-_T Z_T$ (blue) and $c_\theta = 1$ for $W^-\gamma$ (purple). The distribution of the longitudinal polarizations of $W_LZ_L$ (red)  coincides with the Higgs channel $WH$ (black). The vertical axes on the right-handed side again show the number of events per bin expected at the 10 TeV muon collider. In the clean lepton collider environment, we expect the other SM background to be manageable. It is promising to observe the EWSR at the level of $\delta\approx M_W/2~{\rm TeV} < 5\%$ at the energy illustrated. It is conceivable to improve the sensitivity at higher $M_{WZ}$ as long as a sufficient number of events are reconstructed. 


{\bf Discussions and Summary} - Although the Goldstone bosons are only associated with the spontaneously broken generators, approaching the limit of the GET is the necessary step towards EWSR. We have proposed a quantitative measure of symmetry breaking effects in Eq.~(\ref{eq:d}). Depending on the couplings of the Goldstone bosons in the symmetric phase, this should be made quantitative, as 
\begin{equation}
   \delta \ll g,\ g',\ y_f. 
\end{equation}
For a light fermion $f$, the condition $\delta<y_f$ is hard to satisfy, indicating the degree of the GET violation and the EW symmetry non-restoration with respect to that particular Yukawa coupling $y_f$. A good example of this non-restoration is a light fermion splitting to a longitudinal gauge boson $f\to f W_L$. The leading term is due to the EWSB effect $M_W^2/k_T^2$, resulting in the ``ultra collinear behavior'' \cite{Chen:2016wkt}, which is typically much larger than the contribution of the scalar Yukawa coupling $y_f$ in the symmetric phase. 

Although EWSR can be established by the proposed comparative studies at high energies, we reiterate that we only (experimentally) observe the particles in the broken phase. Interestingly, projecting to the symmetric phase from Eqs~(\ref{eq:wgm}) and (\ref{eq:wzT}), the SU(2) $W_i W_3$ scattering amplitude produces a RAZ at $\cos\theta =0$, while the hypercharge gauge-boson process $W_i B$ does not develop a RAZ due to the Abelian nature of the interaction. However, in the experiment, we only observe $B,W_3 \to \gamma,Z$ in the  detectors. This is analogous to QCD, where high-energy processes can be described by quark/gluon dynamics, but the experimental observation is in the broken phase $q,g\to$ hadronic jets in detectors. Along the line, as already pointed out, the SM particles and high-energy scattering processes should be properly treated in the symmetric phase in the partonic picture, and the EW parton distribution functions \cite{Han:2020uid,Han:2021kes,Garosi:2023bvq} and the parton showering \cite{Chen:2016wkt,Bauer:2018xag,Han:2022laq} should be formulated accordingly, while the experimental observables still manifest themselves as massive particles in the broken phase. 

In summary, we introduced a quantitative measure of electroweak symmetry restoration (EWSR) 
by $\delta=M_W/2E$.
We proposed to study EWSR via three processes for the gauge-boson pair production by utilizing their unique gauge structures of RAZs. The LHC experiments have already established the observation of the $W\gamma,\ WZ$ processes, consistent with the SM prediction. Combined analyses including the $WH$ channel could lead to exploration of EWSR at a level of $\delta \sim 10\%$. In a future high-energy lepton collider, such as a multi-TeV muon collider, the unique processes $\mu^\pm \nu \to W^\pm X$ would be able to further improve the measurement accuracy to a percentage level. This can be complementary to the challenging task of observing the $W_L W_L$ scattering. Ultimately, EWSR would be best tested with a high temperature effective potential at $T\sim v\sim 250$ GeV, perhaps only indirectly studied from the early universe cosmology \cite{Weinberg:1974hy}. While testing EWSR is an interesting process in its own right, observing deviation from the SM expectation would be more exciting in the hope of discovering the BSM physics such as an extended Higgs sector, or new strong dynamics associated with EWSB, or providing a verdict for the SMEFT or HEFT formulation.

{\bf Acknowledgments} - We thank Ayres Freitas, Zhen Liu and Michael Peskin for valuable discussions. This manuscript was authored by Fermi Research Alliance, LLC under Contract No.~DE-AC02-07CH11359 with the U.S. Department of Energy, Office of High Energy Physics. TH would like to thank the Aspen Center for Physics which is supported by the National Science Foundation (NSF) grant PHY-1607611, and the CERN TH Department for hospitality during the final stage of the project. This work was supported in part by the U.S.~Department of Energy under grant No.~DE-SC0007914 and in part by Pitt PACC.

\bibliography{Ref}

\end{document}